\documentclass{Interspeech2024}

\interspeechcameraready 

\usepackage[utf8]{inputenc} 
\usepackage[T1]{fontenc}    
\usepackage{hyperref}       
\usepackage{url}            
\usepackage{booktabs}       
\usepackage{amsfonts}       
\usepackage{nicefrac}       
\usepackage{microtype}      
\usepackage{xcolor}         
\usepackage{cite} 
\hypersetup{colorlinks=true}
\usepackage[page]{appendix}
\usepackage{caption}
\usepackage{subcaption}
\usepackage{graphicx}
\usepackage{algorithm}
\usepackage{algpseudocode}
\usepackage{fancybox,framed}
\usepackage{multirow}
\usepackage{tabularx}
\usepackage{soul}
\usepackage{amsmath}
\usepackage{enumitem}
\usepackage{multirow}
\usepackage{diagbox}
\usepackage[export]{adjustbox}
\usepackage{array}

\title{Adversarial training of Keyword Spotting to Minimize TTS Data Overfitting}

\name[affiliation={1}]{Hyun Jin}{Park}
\name[affiliation={1}]{Dhruuv }{Agarwal}
\name[affiliation={1}]{Neng}{Chen}
\name[affiliation={1}]{Rentao}{Sun}
\name[affiliation={1}]{Kurt}{Partridge}
\name[affiliation={1}]{Justin}{Chen}
\name[affiliation={1}]{Harry}{Zhang}
\name[affiliation={1}]{Pai}{Zhu}
\name[affiliation={1}]{Jacob}{Bartel}
\name[affiliation={1}]{Kyle}{Kastner}
\name[affiliation={1}]{Gary}{Wang}
\name[affiliation={1}]{Andrew}{Rosenberg}
\name[affiliation={1}]{Quan}{Wang}

\address{
  $^1$Google LLC, Mountain View, CA, U.S.A.}

\email{\{hjpark, dhruuv, nengchen, sunrentao, kep, jstchen, harryz, paizhu, bartel, kkastner, wgary, rosenberg, quanw\}@google.com}


\begin{document}

\maketitle

\begin{abstract}
    
The keyword spotting (KWS) problem requires large amounts of real speech training data to achieve high accuracy across diverse populations. Utilizing large amounts of text-to-speech (TTS) synthesized data can reduce the cost and time associated with KWS development. However, TTS data may contain artifacts not present in real speech, which the KWS model can exploit (overfit), leading to degraded accuracy on real speech. To address this issue, we propose applying an adversarial training method to prevent the KWS model from learning TTS-specific features when trained on large amounts of TTS data.
Experimental results demonstrate that KWS model accuracy on real speech data can be improved by up to 12\% when adversarial loss is used in addition to the original KWS loss. Surprisingly, we also observed that the adversarial setup improves accuracy by up to 8\%, even when trained solely on TTS and real negative speech data, without any real positive examples.

\end{abstract}

\section{Introduction}
\label{sec:introduction}

Keyword Spotting (KWS) is a task to detect spoken keywords while ignoring background speech and noise. KWS is an important mechanism for virtual assistants to initiate interaction with users via spoken language~\cite{Alvarez2019, AlexaMulti16, HeySiri17}.

A production KWS system needs to detect keywords accurately across diverse populations, acoustic environments, and overlapping noise conditions. Additionally, KWS systems usually need a small footprint to meet the requirements of always-on streaming applications~\cite{small_fp_kws}.

To meet these constraints, neural networks have been extensively studied for KWS. Prior work has demonstrated significant improvements in quality and reductions in latency in low-resource inference settings~\cite{small_fp_kws, HeySiri17, comp_td_nn_kws, MaxPool20, cascade_kws, alvarez2019end, park2020learning}.

Despite numerous technical improvements, production KWS models still require large amounts of data to cover diverse pronunciations and environments. Gathering keyword-specific audio data often involves significant effort and cost, frequently requiring human contributors to generate recordings.

Recent advancements in TTS systems~\cite{Saeki2022VirtuosoMM, Saeki2024ExtendingMS, Borsos2022AudioLMAL} allow for the generation of realistic speech data at scale, which can be used for KWS training. For the same volume of training data, TTS is significantly cheaper and faster than collecting real audio.

However, despite recent advancements in TTS technology, the distribution of generated TTS data may not match that of real data~\cite{Hu2021SYNTUI, Chen2021ContrastiveSG}. In particular, TTS-generated data might lack the diversity present in real human speech and may contain TTS artifacts or other hidden features that can lead to overfitting in machine learning (ML) models.

Such overfitting can make a KWS model less responsive to real positive target speech. This risk is especially high when the amount of real positive data is very small, while a large amount of synthetic data is used. In such cases, a compensatory mechanism can help prevent models from overfitting to the synthetic data.

Adversarial techniques have been previously applied to reduce overfitting to specific domain data and improve generalization to novel domains~\cite{Ganin2014UnsupervisedDA, Sun2017AnUD, Donahue2017ExploringSE, Meng2018SpeakerInvariantTV, Wang2018UnsupervisedDA, Meng2019AdversarialSA}.
In these approaches, an adversarial classifier is trained to predict or discriminate the domain of the input data based on features and representations from the main task model. The main task model's features and representations are then adapted adversarially to become less sensitive to the input data domain. This approach has been shown to successfully improve the generalization of main task models, making them less dependent on the specific data domain.


In this paper, we explore the use of adversarial training (domain adaptation) techniques for a KWS model trained with large amounts of TTS data. In our setup, we propose adding an adversarial classifier that learns to predict whether an input example is synthetic or real speech based on the KWS model's hidden layer features. The loss from this synthetic/real (S/R) classifier is adversarially applied to the KWS model weights to reduce any information that differentiates TTS from real data.
In our experiments, we first show that an adversarial classifier can achieve reasonably high prediction accuracy, demonstrating that the KWS model features do contain TTS-specific features. We then show that accuracy on a real speech evaluation set can be improved by applying adversarial loss to the KWS model under certain data mixture conditions. Surprisingly, adversarial training can also improve model accuracy even when trained solely on real negative data, without any real positive data.

\section{Related work}
\label{sec:relatedworks}

With the advancement of TTS technology, there have been works to explore using TTS data on KWS model development~\cite{Werchniak2021ExploringTA, Lin2020TrainingKS, Lee2022KeywordSW, Kesavaraj2024OpenVK}. These works showed some success in low resource scenarios.

Prior work in automatic speech recognition (ASR) and image processing has noted the potential mismatch between synthetic and real audio data and sought to reduce this discrepancy.
Hu et al.~\cite{Hu2021SYNTUI} discussed the gap between real audio data and TTS synthesized data distributions, categorizing TTS-generated samples as over-sampled, under-sampled, missing, or artifact-containing, depending on the relative likelihood in the real speech data distribution. Artifact regions denote cases where TTS-generated data has novel features not seen in real audio distributions, while missing regions denote the opposite. Both "missing" and "artifact" TTS-generated data can lead to overfitting in machine learning models, hindering generalization to real speech data. Hu et al.~\cite{Hu2021SYNTUI} attempted to address this issue through controlled sampling and separate normalization of synthetic and real data.

In the image classification domain, Chen et al.~\cite{Chen2020AutomatedSG, Chen2021ContrastiveSG} addressed the generalization problem when synthetic data is used to train a recognizer for new classes of objects. The authors used a real-data trained ImageNet model as a teacher and applied transfer learning techniques to improve generalization to real images.

The adversarial training technique was first introduced in a generative modeling context~\cite{Goodfellow2014GenerativeAN}, but it has also been applied to domain mismatch and adaptation problems~\cite{Ganin2014UnsupervisedDA, Sun2017AnUD, Donahue2017ExploringSE, Meng2018SpeakerInvariantTV, Wang2018UnsupervisedDA, Meng2019AdversarialSA}. In this context, adversarial training encourages a neural network to learn features (or representations) that are invariant across different conditions such as environment, noise levels, and speakers.

Motivated by prior works, we apply adversarial training to reduce the representation mismatch between synthetic TTS and real speech domains, in the context of utilizing TTS-generated data for KWS. Specifically, we use adversarial loss to align the hidden representations of the KWS model so that it can generalize better to real speech data. To our knowledge, our approach is the first to use adversarial training to prevent overfitting to synthetic data in KWS and the speech processing domain.

\section{Baseline keyword spotting model}
\label{sec:kws_baseline}

\subsection{Input features}

For the input features, we adopted the same configuration used in prior publications~\cite{Alvarez2019,MaxPool20}. A 40-dimensional vector representing spectral filter-bank energies over a 25-millisecond window is computed every 10 milliseconds. We stacked three temporally adjacent frames, striding by two, to produce a 120-dimensional input feature vector $X_t$ every 20ms. To improve model robustness and generalization, we applied data augmentation~\cite{kim2017generation}, as in prior work.

\subsection{Architecture} 
\label{sec:architecture}

\begin{figure}
	\centering
	\includegraphics[width=\columnwidth]{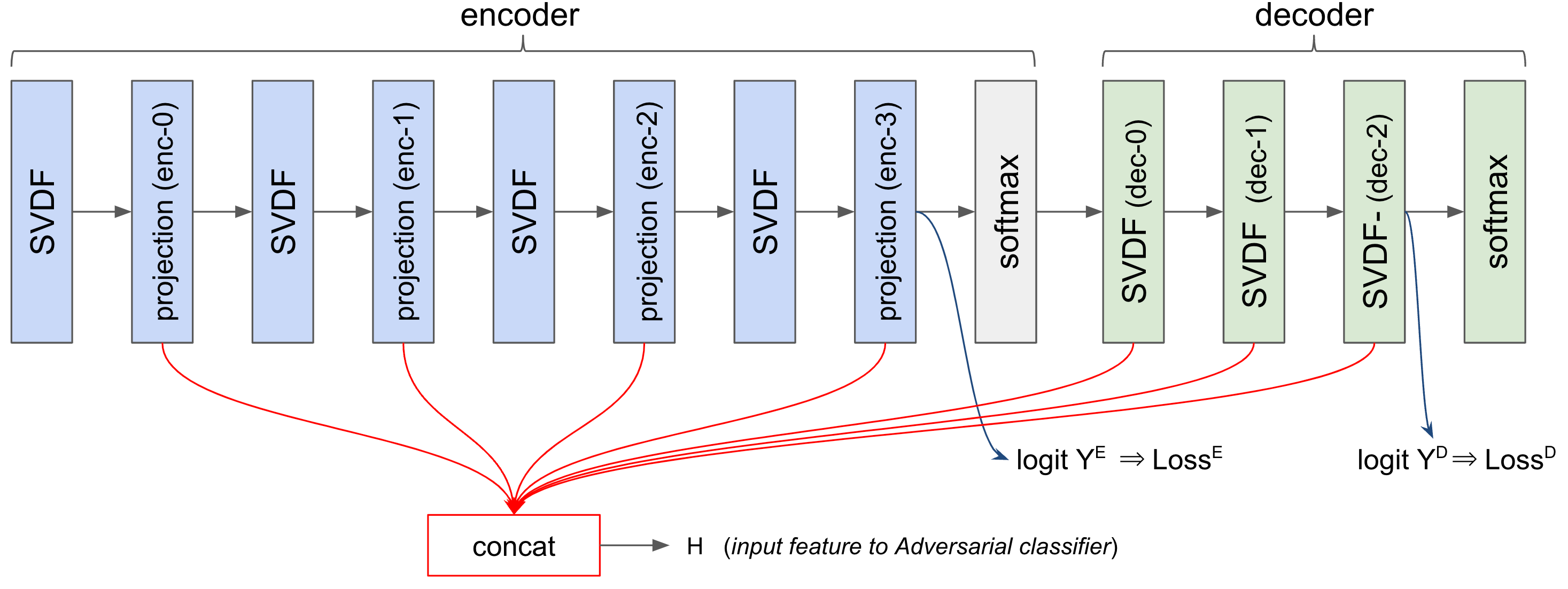}
	\caption{Baseline KWS model architecture and input feature for adversarial classifier}
	\label{fig:baseline_and_hidden_layer_feature}
\end{figure}

Following prior publications~\cite{Alvarez2019, MaxPool20} , we adopted the two-stage model architecture  (Fig.~\ref{fig:baseline_and_hidden_layer_feature}) for both the baseline and proposed approaches.
The KWS model consists of seven factored convolution layers (called SVDF
\cite{Alvarez2019}) and three bottleneck projection layers, organized into sequentially connected encoder and decoder sub-modules. The model contains approximately 320,000 parameters in total.
The encoder module takes as input the feature vector $X_t$, which is a stack of spectral filter-bank energies. It generates a $K$-dimensional output $Y^\textrm{E}$, trained to encode $K$ phoneme-like sounds using ASR-aligned phoneme targets. The decoder module processes the encoder output and generates a 2-dimensional output $Y^\textrm{D}$ trained to predict the existence of a keyword in the input audio stream. The combined prediction logit is defined as $Y=[Y^\textrm{E},Y^\textrm{D}]$. 

\subsection{Training objective}
\label{sec:supervised}

The baseline KWS model is trained by two types of supervised losses.
The first loss computes cross-entropy between model logits and
labels~\cite{Alvarez2019}. The second loss term computes the cross-entropy between max-pooled logits and labels~\cite{MaxPool20}. Both loss terms have separate components for the encoder and decoder, and a weighted combination of all terms is used in the final loss (Eq.~\ref{eq:mlmp-loss}).

\begin{equation}
\label{eq:mlmp-loss}
\begin{split}
\mathcal{L}_{\textrm{sup}} = \sum_{t=1..n} [&(1 - \alpha) L_{\textrm{CE}}\left (Y(X_t,\theta), c_t  \right ) \\
 &+ \alpha L_{\textrm{MP}}\left (Y(X_t,\theta), \omega_{\textrm{end}}  \right )]
\end{split}
\end{equation}

$Y(X_t,\theta)$ represents the combined encoder and decoder model output given input $X_t$ and parameter set $\theta$. $L_{\textrm{CE}}$ represents the end-to-end cross-entropy loss proposed by Alvarez et al.~\cite{Alvarez2019} and $\alpha$ is an weighting term. The implementation from Eq. 2 in Park et al.~\cite{MaxPool20} was used, where $c_t$ is the per-frame target
label for cross-entropy loss. $L_{\textrm{MP}}$ represents the max-pool loss from Eq. 12 in Park et al.~\cite{MaxPool20}.
$\omega_{\textrm{end}}$ represents the end-of-keyword position label for the max-pool loss (refer to Fig.2 in \cite{MaxPool20}). $\alpha$ is an empirically-determined loss-weighting hyper-parameter.

\section{Proposed approach}
\label{sec:proposed_method}

In our proposed approach, the baseline KWS model is augmented with an adversarial classifier that takes hidden layer activations from the baseline model and predicts whether the input data is from a synthetic source (TTS) or real human speech recordings.

We insert a gradient reversal layer \cite{Ganin2014UnsupervisedDA} between the adversarial classifier and the KWS model. This allows the adversarial classifier to adapt and minimize the synthetic/real (S/R) classification loss while simultaneously adapting the KWS model weights to increase the S/R classifier loss. This adversarial gradient update modifies the KWS model weights, making it more difficult for the S/R classifier to discriminate between synthetic and real input data. Concurrently, we minimize the conventional KWS loss to adapt the KWS weights for better keyword detection accuracy.

\begin{figure}
	\centering
	\includegraphics[width=\columnwidth]{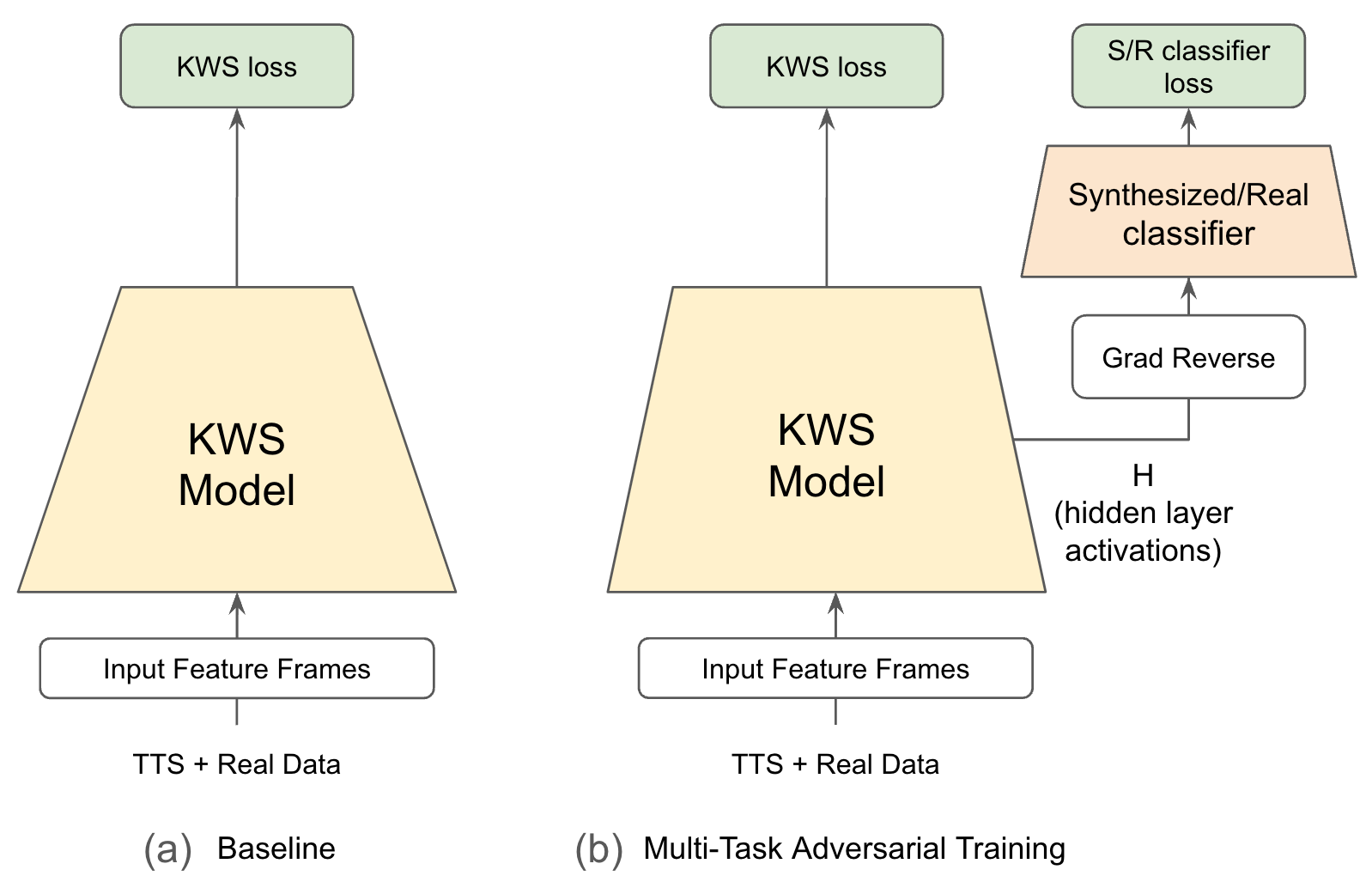}
	\caption{Baseline vs Proposed adversarial training method.}
	\label{fig:architecture}
\end{figure}

The loss function for the proposed approach is shown in Eqs.~\ref{eq:proposed-loss}, \ref{eq:adversarial-loss}, and \ref{eq:adversarial-classifier}. We combine the KWS loss $\mathcal{L}_{\textrm{sup}}$ with the adversarial loss $\mathcal{L}_{\textrm{adv}}$ in a multi-task learning framework.

\begin{equation}
\label{eq:proposed-loss}
\mathcal{L}_{\textrm{total}} = (1 - \beta) \cdot \mathcal{L}_{\textrm{sup}} + \beta \cdot \mathcal{L}_{\textrm{adv}}
\end{equation}

\begin{equation}
\label{eq:adversarial-loss}
\mathcal{L}_{\textrm{adv}} = GR(L_{\textrm{CE}}\left (Y_{adv}(H ; \theta_{adv}), C_{adv}  \right))
\end{equation}

\begin{equation}
\label{eq:adversarial-classifier}
Y_{adv}(H ; \theta_{adv}) = Maxpool(W_{adv} * H_t)
\end{equation}

$Y_{adv}$ is the output of the adversarial classifier, shown as the Synthetic/Real classifier in Fig.~\ref{fig:architecture}. The classifier takes as input $H=[H_t]_{t=0...n}$ (the full sequence of hidden layer activations from the KWS model), and generates a binary classification output predicting whether the input is from a synthetic TTS source or real human speech.
$H_t$ is the hidden layer feature vector at frame $t$ as shown in Fig. \ref{fig:baseline_and_hidden_layer_feature}. $C_{adv}$ is the label for the synthetic/real classifier.
The gradient reversal layer \cite{Ganin2014UnsupervisedDA} is inserted between input $H$ and the synthetic/real classifier to train the KWS model weights. We also use a gradient scaling factor ($\lambda$), which scales the gradient back-propagated from the adversarial classifier into the KWS model.

Various neural network models, such as transformers or LSTMs, can be used to compute $Y_{adv}$. In our implementation, we applied linear projection at each frame, followed by a max-pooling operation over time, to produce a binary logit (Eq. \ref{eq:adversarial-classifier}). We found this method achieves high classification accuracy (up to 98\%), indicating that there are relatively simple features in the audio that can differentiate real from synthetic audio.

\section{Experimental setup}
\label{sec:experiments}

\subsection{TTS system}
We generated TTS data using two systems: Virtuoso~\cite{Saeki2022VirtuosoMM, Saeki2024ExtendingMS} and a variant of the AudioLM \cite{Borsos2022AudioLMAL}. Both are capable of generating data for hundreds of synthetic voices across dozens of locales.
Virtuoso is a multilingual speech-text joint training model that can learn from untranscribed speech, unspoken text, and paired speech-text data sources. AudioLM is a language-model-based audio generation model that features long-term coherence and high quality. We used a variant of the AudioLM model that can be conditioned on both text and sample audio.

\subsection{Dataset}

We compared the baseline and proposed approaches on the "Hey/OK Google" target keyword detection task. For real speech data, we used anonymized utterances collected in accordance with Google's Privacy and AI Principles \cite{privacyprinciples, aiprinciples}. TTS data were generated using Virtuoso and a variant of the AudioLM TTS model (equally sampled). Multi-style data augmentation \cite{KimMTR2017} was applied during training. Table \ref{tab:training_data} summarizes the number of utterances used.

\begin{table}[h!]
	\begin{center}
		\caption{Data types and sizes}
		\label{tab:training_data}
		\begin{tabular}{l|r}

		\hline
		Data Types &  Utterance counts  \\
		\hline
Real Positive Utts &		3.8 M \hspace{0.6cm} \\
Real Negative Utts &		14.1 M \hspace{0.6cm}  \\
		\hline
Synthesized Positive Utts 	&	7.5 M \hspace{0.6cm}  \\
Synthesized Negative Utts	&	5.1 M \hspace{0.6cm}  \\
		\hline

		\end{tabular}
	\end{center}
\end{table}

We tried to use a relatively small amount of real positive utterances, while maximizing the use of real negative audio. This is because we can utilize negative audio with any content, as long as it does not contain the target keywords. Conversely, real positive data is generally not very common in typical data sources and requires costly acquisition processes.

We also used relatively large amounts of TTS synthesized data for both positive and negative conditions, as TTS data is cheaper to produce.

TTS positive data was generated from transcripts sampled from ASR transcripts of real positive utterances. We inserted the target keyword into the sample transcripts to mimic natural positive utterances. TTS voice types were randomly sampled. We also randomly added prosody control symbols to the transcripts. For example, Virtuoso TTS supports controls such as "pause" and "speak slowly" by inserting special characters into the transcript.

\subsection{Feature selection for adversarial classifier}

To determine which features to feed into the Synthetic/Real classifier, we conducted a sweep with various options to select different hidden layers from the KWS model. We used a concatenation operation to combine multiple hidden layer activations. We then trained model (b) from Fig~\ref{fig:architecture} with the gradient reversal operation replaced by a gradient stop operation, so that the KWS model would not be affected by the adversarial classifier. We then evaluated the prediction accuracy of the adversarial classifier for each input option (Table~\ref{tab:feature_selection}).

\begin{table}[h!]
	\begin{center}
		\caption{Adversarial classifier accuracy vs Input features}
		\label{tab:feature_selection}
		\begin{tabular}{c|l}
			\hline
			Adversarial accuracy &  Input features \\
			\hline
 98.1 \% & $en_0$, $en_1$, $en_2$, $en_3$, $de_0$, $de_1$, $de_2$\\
 97.8 \% & $en_0$, $en_1$, $en_2$, $en_3$  \\
 97.1 \% & $en_0$, $en_1$, $en_2$  \\
 96.1 \% &  $en_0$, $en_1$   \\
 96.0 \% & $en_2$   \\
 95.5 \% & $en_3$    \\
 95.3 \% & $en_1$    \\
 92.9 \% &  $de_0$, $de_1$, $de_2$  \\
 91.2 \% &  $de_0$  \\
 89.7 \% & $en_0$   \\
 89.7 \% & $de_1$  \\
 87.1 \% & $de_2$  \\
			\hline

		\end{tabular}
	\end{center}
\end{table}

Table \ref{tab:feature_selection} shows a list of input feature combinations ordered by prediction accuracy. It reveals that concatenating all hidden layer activations yields the best prediction accuracy for the Synthetic/Real classifier. Based on this result, we chose to use all hidden layer activations as input for the subsequent adversarial training experiments. We also observed that the prediction accuracy is generally high, indicating that the KWS model's hidden layer activations and input audio contain significant information that differentiates real from synthetic audio.

\subsection{Adversarial Training setup}
\label{sec:adv_training_setup}

We trained the baseline and proposed adversarial training methods, as shown in Fig. \ref{fig:architecture}, while varying hyperparameters such as data weighting and the gradient reversal scaling factor ($\lambda$).

We varied the sampling probability of the real positive data between 0\% and 100\% to simulate various conditions with different amounts of available real positive data. For example, a 0\% sampling probability corresponds to the case where no real positive data is available, while 100\% corresponds to the case where the full 3.8 million utterances from real positive data can be used.

In preliminary experiments, we found that the model's performance and convergence depended on the choice of gradient scaling factor $\lambda$ applied at the gradient reversal layer, so we swept over multiple $\lambda$ values.

\section{Results}
\label{sec:results}

Table~\ref{tab:exp_result} summarizes the results of training the baseline and proposed approaches under various sweep conditions.

\begin{figure}
	\centering
	\includegraphics[width=\columnwidth]{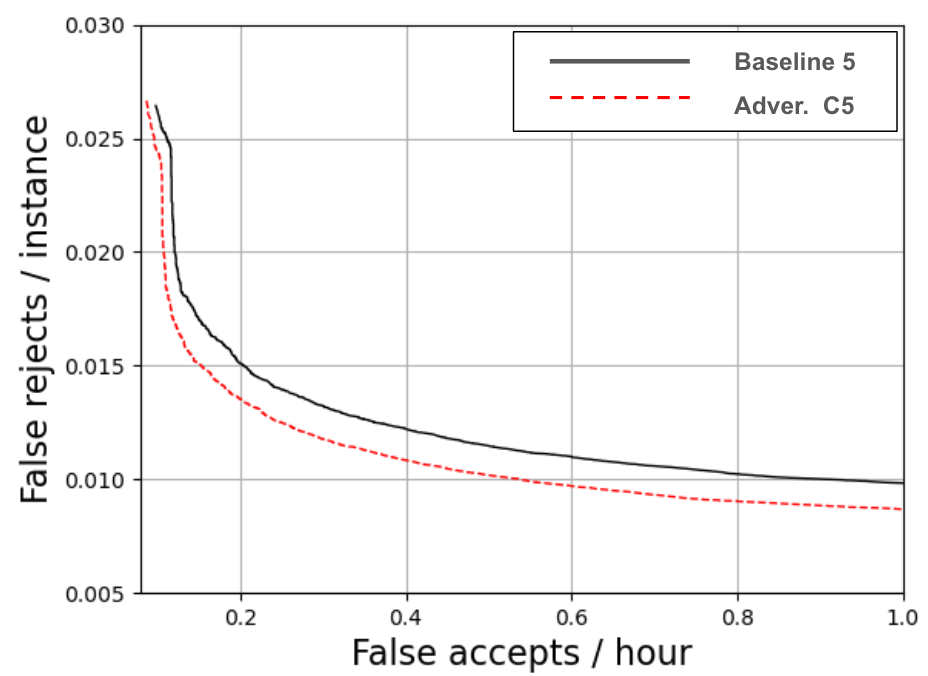}
	\caption{ROC plot of Baseline 5 vs Adversarial Sweep C5}
	\label{fig:roc_plot}
\end{figure}

The FRR (False Rejection Rate) numbers in Table~\ref{tab:exp_result} are taken at thresholds that yield a fixed FA/h (false accept per hour) level on the negative evaluation set. We targeted 0.133 FA/h as the fixed level for all FRR calculations. Figure~\ref{fig:roc_plot} shows the ROC (Receiver Operating Characteristic) plot for the best-performing model (Adversarial C5) compared to the baseline, and it indicates that the relative improvements are consistent across a wide range of FA/h values. This supports the use of FRR at a fixed FA/h level as a representative metric.

$\lambda$ is the gradient scaling factor at the gradient reversal layer, determining how strongly the adversarial gradient affects the KWS model weights. We tested a range of values and report the range that worked reasonably well ($\lambda$= 0.30$\sim$0.50).

Real Positive data Weights (R.Pos.W.) determine the proportion of real positive data examples sampled during training. We swept between 0\% and 100\% as mentioned in section \ref{sec:adv_training_setup}.

The first five rows in the Table~\ref{tab:exp_result} correspond to the baseline model trained with different real positive data weights. The next 20 rows correspond to the proposed adversarial model with different $\lambda$ and real positive data weights. The last five rows are averaged across $\lambda$ values.

\begin{table}[h!]
	\begin{center}
		\caption{KWS model accuracy over hyper parameter options. Table columns include sweep conditions ($\lambda$, and Real Positive data Weights), the FRR (false rejection rate) on the keyword spotting eval set, and relative improvement (R. Imp.) of FRR numbers compared to the baseline models.
		}
		\label{tab:exp_result}
		\begin{tabular}{c|r|r|r|r}
			\hline
			Model & $\lambda$ & R.Pos.W. & Kws FRR & R.Imp. \\
			\hline
			Baseline 1 & - & 0\% & 18.11\% & -  \\ 
			Baseline 2 & - & 1\% & 6.83\%  & -  \\ 
			Baseline 3 & - & 5\% & 3.51\%  & -  \\ 
			Baseline 4 & - & 20\% & 2.38\%  & -  \\ 
			Baseline 5 & - & 100\% & 1.81\%  & -  \\ 
			\hline
			\hline
			Adv. A1  & 0.30 & 0\% & 16.60\% & \textbf{8.3}\%  \\ 
			Adv. A2  & 0.30 & 1\% & 6.87\% & -0.6\%  \\ 
			Adv. A3  & 0.30 & 5\% & 3.58\% & -2.0\%  \\ 
			Adv. A4  & 0.30 & 20\% & 2.12\% & 10.9\%  \\ 
			Adv. A5  & 0.30 & 100\% & 1.61\% & 11.0\%  \\ 
			\hline
			Adv. B1  & 0.35 & 0\% & 16.89\% & 6.7\%  \\ 
			Adv. B2  & 0.35 & 1\% & 6.52\% & 4.5\%  \\ 
			Adv. B3  & 0.35 & 5\% & 3.87\% & -10.3\%  \\ 
			Adv. B4  & 0.35 & 20\% & 2.28\% & 4.2\%  \\ 
			Adv. B5  & 0.35 & 100\% & 1.65\% & 8.8\%  \\ 
			\hline
			Adv. C1  & 0.40 & 0\% & 16.84\% & 7.0\%  \\ 
			Adv. C2  & 0.40 & 1\% & 6.67\% & 2.3\%  \\ 
			Adv. C3  & 0.40 & 5\% & 4.04\% & -15.1\%  \\ 
			Adv. C4  & 0.40 & 20\% & 2.40\% & 0.8\%  \\ 
			Adv. C5  & 0.40 & 100\% & 1.59\% & \textbf{12.2\%}  \\ 
			\hline
			Adv. D1  & 0.50 & 0\% & 17.62\% & 2.7\%  \\ 
			Adv. D2  & 0.50 & 1\% & 6.45\% & 5.6\%  \\ 
			Adv. D3  & 0.50 & 5\% & 3.58\% & -2.0\%  \\ 
			Adv. D4  & 0.50 & 20\% & 2.25\% & 5.5\%  \\ 
			Adv. D5  & 0.50 & 100\% & 1.62\% & 10.5\%  \\ 
			\hline
			\hline
			Averaged 1  & 0.3-0.5 & 0\% & 16.99\% & 6.2\%  \\ 
			Averaged 2  & 0.3-0.5 & 1\% & 6.63\% & 3.6\%  \\ 
			Averaged 3  & 0.3-0.5 & 5\% & 3.77\% & -7.3\%  \\ 
			Averaged 4  & 0.3-0.5 & 20\% & 2.26\% & 4.9\%  \\ 
			Averaged 5  & 0.3-0.5 & 100\% & 1.62\% & 10.6\%  \\ 
			\hline
		\end{tabular}
	\end{center}
\end{table}

The results in Table~\ref{tab:exp_result} show that adversarial training can improve KWS model accuracy when either real positive data weights are very low (near 0\%) or very high (near 100\%). Relative improvements are strongest when using the full amount of real positive data (average 10.6\%). Surprisingly, adversarial training can also improve KWS model accuracy (average 6\%), even without any real positive data. Note that we still include \textbf{real negative data}, which can serve as contrasting examples against synthetic data to train the adversarial classifier. We observe no improvement or degradation when there is an intermediate amount of real positive data.

\section{Conclusions}
\label{sec:conclusions}

We proposed using adversarial training for keyword spotting (KWS) when large amounts of TTS synthesized data are used. The proposed method builds an adversarial classifier that predicts whether the input source is synthetic or real speech. Its gradients are then adversarially applied to the KWS model, preventing the KWS model from learning features specific to synthesized data.

Results show that when the KWS model is trained with both KWS loss and adversarial loss, its accuracy on real speech data improves by up to 12\% when the full amount of real and TTS data is used. Surprisingly, we also observed that accuracy on real data can be improved (up to 8\%), even when the model is trained solely on TTS and real negative speech data, \textbf{without any real positive examples}. This can be explained by the fact that real negative data can still serve as contrasting examples against synthetic data for the adversarial classifier.

As a further study, it would be interesting to use a more powerful model, such as Conformer or LSTM, for the adversarial classifier.

\section{Acknowledgements}
\label{sec:acknowledgements}

The authors would like to acknowledge the support from Charles Yoon, Pedro Meningbar, Bhuvana Ramabhadran, and Fran{\c{c}}oise~Beaufays.

\pagebreak

\bibliographystyle{IEEEtran}
\bibliography{mybib}

\end{document}